%% file: finalv1.tex
\documentclass[letterpaper, 10 pt, conference]{ieeeconf}  

\IEEEoverridecommandlockouts                              
\usepackage{mathtools} 
\usepackage{fixltx2e} 

\title{\LARGE \bf
Experimental design trade-offs for gene regulatory network inference:\\ an \emph{in silico} study of the yeast \SacCer\,cell cycle
}

\author{Johan Markdahl, Nicolo Colombo, Johan Thunberg, and Jorge Gon\c{c}alves
\thanks{J. Markdahl, J. Thunberg, and J. Gon\c{c}alves are with the Luxembourg Centre for Systems Biomedicine (\textsc{lcsb}),
        University of Luxembourg. Corresponding author:
        {\tt\small jorge.goncalves@uni.lu}}%
\thanks{N. Colombo is with the Department of Statistical Science, University College London (\textsc{ucl}).}%
}

\input{myIncludes}

\begin{document}

\maketitle
\thispagestyle{empty}
\pagestyle{empty}

\begin{abstract}
Time-series of high throughput gene sequencing data intended for gene regulatory network (\GRN) inference are often short due to the high costs of sampling cell systems. Moreover, experimentalists lack a set of quantitative guidelines that prescribe the minimal number of samples required to infer a reliable \GRN{} model. We study the temporal resolution of data \vs quality of \GRN{} inference in order to ultimately overcome this deficit. The evolution of a Markovian jump process model for the \Ras/\cAMP/\PKA{} pathway of proteins and metabolites in the G$_1$ phase of the \SacCer{} cell cycle is sampled at a number of different rates. For each time-series we infer a linear regression model of the \GRN{} using the \LASSO{} method. The inferred network topology is evaluated in terms of the area under the precision-recall curve (\AUPR). By plotting the \AUPR{} against the number of samples, we show that the trade-off has a, roughly speaking, sigmoid shape. An optimal number of samples corresponds to values on the ridge of the sigmoid.
\end{abstract}


\section{Introduction}



\noindent Time-series gene expression data provides a series of snapshots of molecular concentrations in gene regulatory networks (\GRN) \cite{bar2012studying}. This information is used to infer dynamic models of \GRN{} networks which aid our understanding of how observable phenotypes, \eg diseases, arise from molecular interactions \cite{kitano2002systems}. As such, time-series data is of importance to fundamental research within systems biology, and potentially also in applications like medical diagnostics, drug development, and therapies \cite{barabasi2004network}. The advent of high throughput sequencing have made time-series data widely available although it is prohibitively expensive to densely sample gene expression levels. It remains difficult for experimentalists to accurately judge the frequency and distribution of samples needed to infer network structures: for each project, they must navigate the trade-off between oversampling (more samples than necessary, increasing costs with no benefit to \GRN{} inference) and undersampling (too few samples to reliably infer the \GRN, potential waste of resources and failure to infer the \GRN) \cite{bar2004analyzing}. Such costs add up; studies indicate that 85\% of research investment in biomedical sciences is wasted, corrsponding to US\$200  billion worldwide in 2010 \cite{macleod2014biomedical}. This work undertakes an \emph{in silico} study of the impact of the cost \vs number of samples trade-offs on the quality of the output produced by a \GRN{} inference algorithm. Our ultimate goal, to which this paper is a stepping stone, is to formulate guidelines and construct decision support systems to help researches navigate trade-offs such that \GRN{} models of desired quality can be inferred at a minimal cost.




The performance of \GRN{} inference algorithms has been benchmarked against \emph{in silico} and \emph{in vivo} data in a number of comparative studies \cite{werhli2006comparative,marbach2010revealing,marbach2012wisdom,aderhold2014statistical}. The aforementioned trade-off has received comparatively less attention \cite{bar2012studying,sima2009inference,bar2004analyzing,sefer2016tradeoffs,mombaerts2016optimising}. There are of course many works that touch upon it in passing, \eg \cite{husmeier2003sensitivity}, or that pay the price of intentionally oversampling to ensure capturing high-frequency content \cite{owens2016measuring,brunton2016discovering}. 
Early work that take a systematic approach to studying the trade-off are rather abstracts and deal with generalities in broad strokes \cite{bar2012studying,sima2009inference,bar2004analyzing}. For example, \cite{bar2012studying} states that cyclic processes such as cell cycles and circadean rythms should be sampled uniformly over multiple cycles. In perturbation-response studies, by contrast, most samples should be taken early to capture the transient dynamics. 

Only in the past year have results been published to support the common sense notions of navigating the trade-off that are current experimental practices \cite{sefer2016tradeoffs,mombaerts2016optimising}. Sefer \etal \cite{sefer2016tradeoffs} take an in-depth look at the experimental design question of sampling densely versus sampling repeatedly; the former is recommended for the purpose of detecting a spike in the molecule count number of some species. Mombaerts \etal \cite{mombaerts2016optimising} study the difference between transient and steady-state sampling of the circadian clock in \emph{Arabidopsis thaliana}, finding that the transient contains more information. In a similar vein, this paper establishes that the performance of an inference algorithm that fits a linear model to a pathway in the G$_1$ phase of the \SacCer{} cell cycle is comparable to random classifier in the case of 3--6 samples, increases over 7--11 samples, and then flattens out with additional samples giving diminishing returns. Together with \cite{sefer2016tradeoffs}, \cite{mombaerts2016optimising}, this paper represents a first effort to refine previous, rule based experiment trade-off navigation practices \cite{bar2012studying,sima2009inference,bar2004analyzing}, into  more specific, quantitative guidelines.


Alongside the development of novel \GRN{} inference algorithms, new models have been adopted to generate \emph{in silico} data and represent the dynamics of inferred networks \cite{milo2002network,milo2004superfamilies,tyson2003sniffers,wilkinson2009stochastic,karlebach2008modelling}. \GRN{} models exist at different levels of abstraction, from the logical models captured by Boolean networks, over continuous models, \eg systems of ordinary differential equations, to the mesoscopic single molecule models such as chemical reaction networks (\CRN) whose dynamics are modeled as Markovian jump processes governed by the chemical master equation (\CME) \cite{karlebach2008modelling}. To measure the performance of a \GRN{} inference algorithm, the ground truth in terms of gene expression causal interactions is required. For \emph{in vivo} data, the ground truth is often unavailable and replacing it with a known gold standard poses certain challenges \cite{sima2009inference,de2010advantages}, making \emph{in silico} studies an attractive alternative \cite{wilkinson2009stochastic}. In this paper we require \emph{in silico} models to generate output with a wide range of sample rates. We strive to replicate realistic experiment conditions, \eg choosing a detailed \emph{in silico} model of cellular dynamics based on Markovian jump processes to represent key characteristics such as intrinsic noise \cite{wilkinson2009stochastic,mcadams1999sa}, common network motifs like sparsity \cite{milo2002network,milo2004superfamilies}, and species with highly different concentrations \cite{cazzaniga2008modeling}.

This paper uses the \CME{} to model a pathway involved in the G$_1$ phase of the \SCer{} cell cycle \cite{cazzaniga2008modeling}, following the experiment setup of a query driven rather than a global study \cite{de2010advantages}. A sample is drawn from the probability density function governed by the \CME{} using a stochastic simulation algorithm (\SSA). We then infer a linear autoregressive model to explain the \emph{in silico} data using the \LASSO{} method \cite{tropp2010computational}. \LASSO{} provides a basic approach for \GRN{} inference \cite{marbach2012wisdom}, and has the benefit of imposing sparsity on the regression parameters, thereby capturing a characteristic \GRN{} motif. Large regression coefficients suggest the existence of regulatory interactions between species, whereby an interaction topology can be extracted by thresholding the model parameters. The area under the precision-recall curve is used to score the performance of \LASSO{} by comparing the inferred topology with that of the \CRN{} simulated by the \SSA \cite{saito2015precision}. We obtain a graph of the trade-off by repeating the inference procedure for data of varying temporal resolution.  The main contributions of this paper can be summarized as follows: (i) we establish that the trade-off function which charts performance over number of samples has a sigmoid shape for a pathway in the G$_1$ phase of the \SCer{} cell cycle and the \LASSO{} method and (ii) we provide a graph that allows an experimentalist to match a desired quality of inference (for the pathway) with a minimum number of samples.

\section{Research Question and Research Problem}

\label{sec:background}


\noindent Suppose that the experiment budget is somewhat flexible, and that there exist incentives to cut costs. Consider how a biologist conducting a high throughput gene sequencing experiment should navigate the number of samples \vs quality of \GRN{} inference trade-off. Since the cost of undersampling is an incomplete or failed study whereas oversampling amounts to a waste of resources, we express the multiobjective optimization problem, \ie the trade-off, in terms of a hard constraint on the quality of the inferred network: minimize the number of samples required to achieve a certain quality of inference for a given experiment, \ie to optimize marginal costs. For this paper we limit the scope to a particular model of the \Ras/\cAMP/\PKA{} pathway in \SCer{} \cite{cazzaniga2008modeling} and the \LASSO{} method applied to \GRN{} inference \cite{tropp2010computational}. Consider the resolution of gene expressions measurements in cases where additional detail can be purchased at a cost that is higher than that of additional samples, \ie to optimize fixed costs. In particular, we study the cases of including or excluding a phosphoproteomic analysis of \SCer, which requires the use of different techniques compared to proteomics and metabolomics \cite{Larsen2008} (the low molecule count numbers for phosphorylated proteins requires a larger cell culture).


\section{Method}

\label{sec:method}

\noindent To begin with, \emph{in silico} data is generated from a Markov process model of a pathway in the yeast \SCer{} cell cycle, see Section \ref{sec:realistic}. To simulate the model, an efficient solver for the chemical master equation is required as detailed in Section \ref{sec:chem}. The model of the pathway is from \cite{cazzaniga2008modeling}, and has been verified against experimental data. The model consists of molecule count numbers for a total of 30 proteins and metabolites and 34 stochastic reactions. It is described in detail in Section \ref{sec:realistic}. The output of the simulation is sampled at discrete time-points, whereby a sparse discrete-time state-space model is fitted using the \LASSO{} method, see Section \ref{sec:lasso}. The translation of the ground truth causal relations from the Markovian jump process model to a discrete-time difference equation based model is done in Section \ref{sec:causal}. The evaluation of the model in using precision-recall curves based on the relations established in Section \ref{sec:causal} is explained in Section \ref{sec:ROC}. 

\subsection{The chemical master equation}
\label{sec:chem}


\noindent Consider a chemical reaction network (\CRN) from a mesoscopic, non-deterministic perspective  as detailed in \cite{iglesias2010control}. The system consists of $n$ molecular species $S_1,\ldots,S_n$ contained in a volume $\Omega$. The system is assumed to be well-stirred or spatially homogeneous. Let $\ve{X}(t)=[X_1(t),\ldots,X_n(t)]\mtr\in\N^n$ be a vector whose $i$th element $X_i$ denotes the number of molecules of species $S_i$ at time $t$. The $n$ species interact through $m$ reactions $R_1,\ldots,R_m$ on the form
\begin{align}
R_j:\sum_{l=1}^k n_{j_l} S_{j_l}\smash{\xrightarrow{c_j}}\sum_{l=1}^h m_{j_l} P_{j_l},\label{eq:Rj}
\end{align}
where the left-hand side contain the reactants, the right-hand side the products, and $c_j$ is the stochastic reaction constant. Each reaction $R_i$ defines a transition from some state $\ve[0]{X}\in\N^n$ to $\ve{X}(t)=\ve[0]{X}+\ve[i]{S}$, where $\ve[i]{S}$ is a column of the stoichiometry matrix $\ma{S}=[\ve[1]{S}, \ldots, \ve[m]{S}]$.

To each reaction $R_i$ we associate a function $w_i:\N^n\rightarrow[0,\infty)$ such that $w_i(\ve{X})\diff t$ is the probability that $R_i$ occurs just once in $[t,t+\diff t)$ \cite{iglesias2010control}. These, so called propensity functions, $w_i$ are given by $c_i$ times the number of distinct molecular reactant combinations for reaction $R_i$ found to be present in $\Omega$ at time $t$ \cite{Gillespie76}. More specifically, $w_i=c_i$ if $\emptyset\smash{\xrightarrow{c_j}} P$ and 
\begin{align} 
w_i(X_{j_l},\ldots,X_{j_k})=c_i\prod_{l=1}^k\binom{X_{j_l}}{n_{j_l}},\label{eq:propensity}
\end{align}
if $\sum_{l=1}^k n_{j_l}S_{j_l}\rightarrow P$, where $c_i$ is a stochastic reaction constant, $P$ denotes a sum of chemical products, and $n_{j_l}\in\N$ denote the coefficient of $S_{j_l}$ in $R_i$ as detailed in \eqref{eq:Rj}.

Let $\Prob(\ve{X},t):\N^n\times[0,\infty)\rightarrow[0,1]$ denote the probability that the system is in state $\ve{X}$ at time $t$. The chemical master equation (\CME) is a system of coupled differential-difference equations given by
\begin{align}
	\tag{\textsc{cme}}
	\dot{\Prob}(\ve{X},t)&=\sum_{k=1}^m w_k(\ve{X}-\ve[k]{S}) \Prob(\ve{X}-\ve[k]{S},t)-w_k(\ve{x}) \Prob(\ve{X},t),\label{eq:CME}
\end{align}
one equation for each feasible state $\ve{X}\in\N^n$. Any solution to \eqref{eq:CME} corresponds to a sample from $\Prob(\ve{x},t)$. Exact closed-form solutions to \eqref{eq:CME} can only be obtained under rather restrictive assumptions, wherefore most works focus on exact numerical methods, so-called stochastic simulation algorithms (\SSA{}s), approximate numerical methods, \eg the $\tau$-leap algorithm \cite{gillespie2001approximate,cao2006efficient}, or solving approximations to the \CME{} such as the chemical Langevin equation \cite{iglesias2010control}.
 
Gillespie proposes two Monte Carlo \SSA{}s for exact numerical solution of \eqref{eq:CME}: the first reaction method (\FRM) \cite{Gillespie76} and the direct method (\DM) \cite{Gillespie77}. The methods are equivalent since they give the same probability distributions for the first reaction to occur, and the time until its occurrence. The so-called next reaction method (\NRM) allows for more efficient execution of the first reaction method \cite{gibson2000efficient}. However, \cite{gibson2000efficient} underestimated the complexity of the \NRM{} by omitting the cost of managing a priority queue of reaction times \cite{cao2004efficient}. An optimized version of the \DM{} (\ODM) turns out to be more efficient than the \NRM \cite{cao2004efficient}. Additional \SSA{}s have been proposed since then. This paper utilizes the \ODM.

\subsection{The \Ras/\cAMP/\PKA{} pathway in \emph{\SCer}}

\label{sec:realistic}

\noindent The \Ras/\cAMP/\PKA{} pathway is involed in the regulation of  \emph{S. cerevisiae} metabolism and cell cycle progression. A realistic \CRN\,model of 30 proteins and metabolites undergoing 34 reactions is proposed by Cazzaniga \etal \cite{cazzaniga2008modeling}, \cite{besozzi2012role}, see Table \ref{tab:model}. See \cite{williamson2009deterministic} for a deterministic \ODE{} model of the pathway. The pathway is regulated by several control mechanisms, such as the as the feedback cycle ruled by the activity of phosphodiesterase. Feedback and feedforward, \ie directed loops, are common network motifs which pose challanges for many \GRN{} inference algorithms \cite{marbach2010revealing,marbach2012wisdom}. The notation \textbullet{} in Table \ref{tab:model} indicates that two molecules are chemically bound and form a complex. Each complex is treated as a separate variable. For example \GDP, \CdcTF, \RasT\mytextbullet {}\GDP{} and \RasT\mytextbullet {}\GDP\mytextbullet {}\CdcTF{} are four separate variables, three of which appear in reaction one. \RasT{} is however not a variable in this model, as it only appears as part of complexes. The superindex p indicates that a protein is phosphorylated \cite{Larsen2008}. Note that one effect of the chain of reactions $R_1$--$R_{34}$ in Table \ref{tab:model} is to phosphorylate \CdcTF.

\begin{table}
	\begin{center}
		\caption{Stochastic model of the \textsc{r}as/c\textsc{amp}/\textsc{pka} pathway \cite{cazzaniga2008modeling}. Each row of the table represents a reaction on the form of \eqref{eq:Rj}. \label{tab:model}}
		\begin{tabular}[h!]{l c c r}
			\rlap{Reaction} & Reactants  & Products & \llap{Constant}\Bstrut\\
			\hline
			\Tstrut
			$R_1$\phantom{0} & \sRasT\mytextbullet \sGDP{} + \sCdcTF & \sRasT\mytextbullet \sGDP\mytextbullet \sCdcTF & 1e\,0\\
			$R_2$ & \sRasT\mytextbullet \sGDP\mytextbullet \sCdcTF & \sRasT\mytextbullet \sGDP{} + \sCdcTF & 1e\,0\\
			$R_3$\phantom{0} & \sRasT\mytextbullet \sGDP\mytextbullet \sCdcTF & \sRasT\mytextbullet \sCdcTF{} + \sGDP & 1.5e\,0\\
			$R_4$\phantom{0} & \sRasT\mytextbullet \sCdcTF{} + \sGDP & \sRasT\mytextbullet \sGDP\mytextbullet \sCdcTF & 1e\,0\\
			$R_5$\phantom{0} & \sRasT\mytextbullet \sCdcTF{} + \sGTP & \sRasT\mytextbullet \sGTP\mytextbullet \sCdcTF & 1e\,0\\
			$R_6$\phantom{0} & \sRasT\mytextbullet \sGTP\mytextbullet \sCdcTF & \sRasT\mytextbullet \sCdcTF{} + \sGTP & 1e\,0\\
			$R_7$\phantom{0} & \sRasT\mytextbullet \sGTP\mytextbullet \sCdcTF & \sRasT\mytextbullet \sGTP{} + \sCdcTF & 1e\,0\\
			$R_8$\phantom{0} & \sRasT\mytextbullet \sGTP{} + \sCdcTF & \sRasT\mytextbullet \sGTP\mytextbullet \sCdcTF & 1e\,0\\
			$R_9$\phantom{0} & \sRasT\mytextbullet \sGTP{} + \sIraT & \sRasT\mytextbullet \sGTP\mytextbullet \sIraT & 3e-2\\
			$R_{10}$ & \sRasT\mytextbullet \sGTP\mytextbullet \sIraT & \sRasT\mytextbullet \sGDP{} + \sIraT & 7e-1\\
			$R_{11}$ & \sRasT\mytextbullet \sGTP{} + \sCYRO & \sRasT\mytextbullet \sGTP\mytextbullet \sCYRO & 1e-3\\
			$R_{12}$ & \sRasT\mytextbullet \sGTP\mytextbullet \sCYRO{} + \sATP & \sRasT\mytextbullet \sGTP\mytextbullet \sCYRO{} + \scAMP & 1e{-5}\\
			$R_{13}$ & \sRasT\mytextbullet \sGTP\mytextbullet \sCYRO{} + \sIraT & \sRasT\mytextbullet \sGDP{} + \sCYRO{} + \sIraT & 1e-3\\
			$R_{14}$ & \scAMP{} + \sPKA & \scAMP\mytextbullet \sPKA & 1e-5\\
			$R_{15}$ & \scAMP{} + \scAMP \mytextbullet \sPKA & ({\scriptsize2}\scAMP)\mytextbullet \sPKA & 1e-5\\
			$R_{16}$ & \scAMP{} + ({\scriptsize2}\scAMP)\mytextbullet \sPKA & ({\scriptsize3}\scAMP)\mytextbullet \sPKA & 1e-5\\
			$R_{17}$ & \scAMP{} + ({\scriptsize3}\scAMP)\mytextbullet \sPKA & ({\scriptsize4}\scAMP)\mytextbullet \sPKA & 1e-5\\
			$R_{18}$ & ({\scriptsize4}\scAMP)\mytextbullet \sPKA & \scAMP{} + ({\scriptsize3}\scAMP)\mytextbullet \sPKA & 1e-1\\
			$R_{19}$ & ({\scriptsize3}\scAMP)\mytextbullet \sPKA & \scAMP{} + ({\scriptsize2}\scAMP)\mytextbullet \sPKA & 1e-1\\
			$R_{20}$ & ({\scriptsize2}\scAMP)\mytextbullet \sPKA & \scAMP{} + \scAMP\mytextbullet \sPKA & 1e-1\\
			$R_{21}$ & \scAMP\mytextbullet \sPKA & \scAMP{} + \sPKA & 1e-1\\
			$R_{22}$ & ({\scriptsize4}\scAMP)\mytextbullet \sPKA & {\scriptsize2}\sC{} + {\scriptsize2}(\sR\mytextbullet {\scriptsize2}\scAMP) & 1e\,0\\
			$R_{23}$ & \sR\mytextbullet {\scriptsize2}\scAMP & \sR{} + {\scriptsize2}\scAMP & 1e\,0\\
			$R_{24}$ & {\scriptsize2}\sR{} + {\scriptsize2}\sC & \sPKA & 1e\,0\\
			$R_{25}$ & \sC{} + \sPdeO & \sC{} + \sPdeO\shighf & 1e-6\\
			$R_{26}$ & \scAMP{} + \sPdeO\shighf & \scAMP\mytextbullet \sPdeO\shighf & 1e-1\\
			$R_{27}$ & \scAMP\mytextbullet \sPdeO\shighf & \scAMP{} + \sPdeO\shighf & 1e-1\\
			$R_{28}$ & \scAMP\mytextbullet \sPdeO\shighf & \sAMP{} + \sPdeO\shighf & 7.5e\,0\\
			$R_{29}$ & \sPdeO\shighf{} + \sPPAT & \sPdeO{} + \sPPAT & 1e-4\\
			$R_{30}$ & \scAMP{} + \sPdeT & \scAMP\mytextbullet \sPdeT & 1e-4\\
			$R_{31}$ & \scAMP\mytextbullet \sPdeT & \scAMP{} + \sPdeT & 1e\,0\\
			$R_{32}$ & \scAMP\mytextbullet \sPdeT & \sAMP{} + \sPdeT & 1.7e\,0\\
			$R_{33}$ & \sC{} + \sCdcTF & \sC{} + \sCdcTF\shighf & 1e\,1\\
			$R_{34}$ & \sCdcTF\shighf{} + \sPPAT & \sCdcTF{} + \sPPAT & 1e-2\\
		\end{tabular}
	\end{center}
\end{table}

Cazzaniga \etal use the $\tau$-leap algorithm of Gillespie \cite{gillespie2001approximate,cao2006efficient} to solve the \CRN\,model in Table \ref{tab:model} approximately. The stochastic reaction constants in Table \ref{tab:model} have been tuned relatively to each other, but not absolutely wherefore the time-scale of the simulations is given in an unspecified unit \cite{cazzaniga2008modeling}. We prefer to use a known time-scale since the minimum sample time is bounded below for \emph{in vivo} experiments. Experimental results establish that \cAMP{} initially rises to a maximum and then decreases to steady-state with a settling time of 3-5 minutes \cite{rolland2000glucose}. By repeating that experiment \emph{in silico}, \cite{cazzaniga2008modeling} establish that  3--5 minutes correspond to 1000 units of simulation time. The \emph{in vivo} experiment included 15 samples from the evolution of \cAMP{} over 7 minutes \cite{rolland2000glucose}. \textsc{lcsb} experimentalists confirm that we can sample \emph{in vivo} systems at most twice per minute due to technological limitations, corresponding to at most 6--10 samples per 1000 units of simulation time. 



The initial molecule copy numbers from  \cite{cazzaniga2008modeling} are given in Table \ref{tab:copy}. The numbers reflect realistic assumptions regarding the contents of a single cell of \SCer{} based on calculations and experimental data. However, in high throughput gene sequencing experiments, a large number of cells are sampled from a culture and destroyed in the process \cite{alberts1997molecular}. The molecule counts in each sample correspond to a sum of around 50 000 to 100 000 cells. Since any two cells can be in different stages of the \SCer{} cell cycle, their molecule counts may not agree aside from the approximately 10\% difference that is due to intrinsic stochastic variation \cite{alon2006introduction}. This problem is addressed by synchronizing the cell cycles to evolve in phase, for which a number of techniques are available \cite{futcher1999cell}. Under the assumption of \emph{in vivo} data being from a synchronized processes, it is thus justified to study a single cell \emph{in silico}.



\begin{table*}
	\begin{center}
		\caption{Initial values of molecule copy numbers \cite{cazzaniga2008modeling}. Species not listed start at zero molecules.\label{tab:copy}}
		\begin{tabular}[h!]{rccccccccccc}
			Species & \sCYRO & \sCdcTF & \sIraT & \sPdeO & \sPKA & \sPPAT & \sPdeT & \sRasT\mytextbullet {}\GDP & \sGDP & \sGTP & \sATP\\
			 Number & 2e2 & 3e2 & 2e2 & 1.4e3 & 2.5e3 & 4e3 & 6.5e3 & 2e5 & 1.5e6 & 5.0e6 & 2.4e7 
		\end{tabular}
	\end{center}
\end{table*}

\subsection{Network inference method}

\label{sec:lasso}

\noindent \GRN{} inference problems involve many species but few samples and is thus underdetermined \cite{de2010advantages}. A well established network motif, sparsity, \ie that each species interact with only a few other species, is imposed to reduce the number of solutions \cite{alon2006introduction}. Sparsity also protects the inferred model against overfitting without having to deal with the combinatorial explosion that other methods for model selection such as those based on the Akaike or Bayesian information criteria face. A basic problem in compressive sampling, to find the sparsest solution to a linear system of equations in terms of the number of nonzero entries, is \NP-hard \cite{natarajan1995sparse} and difficult to approximate \cite{amaldi1998approximability} wherefore the use of convex relaxations and other heuristic methods is commonplace \cite{tropp2010computational}. A dynamical system is usually not the object of study in compressive sampling \cite{candes2008introduction}, although techniques from that field can be used for \GRN{} inference. To adopt a convex relaxation of the sparse approximation technique to time-series we use the idea of minimizing an error.

To explain the discrete \GRN{} data $\ve{X}(t)\in\N^n$ for all $t\in[0,\infty)$, we adopt a discrete-time system model,
\begin{align*}
\vh[k+1]{X}&=\ve{f}(\Delta t_k,\vh[k]{X})+\ve[k]{\varepsilon},
\end{align*}
where $\vh[k]{X}\in\R^n$ models $\ve{X}(t_k)$, $\Delta t_k=t_{k+1}-t_k$, and  $\ve[k]{\varepsilon}$ is white noise. For the sake of simplicity we take $\ve{f}:\R^{n}\rightarrow\R^n$ to be a linear function, \ie
\begin{align}
\vh[k+1]{X}&=\ma{A}(\Delta t_k)\vh[k]{X}+\ve[k]{\varepsilon}.\label{eq:linear}
\end{align}
Since the propensity functions \eqref{eq:propensity} of the \CME{} are nonlinear, the model \eqref{eq:linear} will not capture all the species interdependencies and we cannot expect a zero error in the limit of infinite samples. However, rather than adding a large dictionary of terms that are linear in parameters but nonlinear in the explanatory variables we prefer to adopt a minimal model. The limit would anyhow not be approached in practice due to the low temporal resolution of data, and there is merit to using linear models since certain nonlinear \GRN{} models are prone to overfitting \cite{aderhold2014statistical}. Since the \Ras/\cAMP/\PKA{} pathway is part of a cell cycle, we take the advice of \cite{bar2012studying} and adopt a uniform sample rate, \ie $\Delta t_k=\Delta t\in(0,\infty)$ in \eqref{eq:linear}. This requires some post-processing of the \SSA{} data.

The output of the \SSA{} consists of the molecule count numbers and time instances for each reaction during a timespan $[0,T]$. To create discrete-time samples $(\ve{X}(t_k))_{k=0}^{N-1}$ with $t_0=0$, $t_k=T$, $t_{i+1}-t_i=\Delta t$, for all $i=0,\ldots,N-1$ we use the \matlab{} function \texttt{interp1} that interpolates linearly based on the data obtained from the \SSA\,and rounds each sample to the nearest point in $\N^n$. The output from the \SSA\,contains a number of time-points on the order of $10^8$ whereas $T$ is on the order of $10^3$, so any error due to the interpolation and rounding is negligible. Since the molecule count numbers vary greatly in order of magnitude, see Table \ref{tab:copy}, we introduce new variables by scaling each time series $(X_i(t_k))_{k=0}^{N-1}$ by a constant equal to one over $\max_{k}X_i(t_k)$ to facilitate the optimization \cite{wright1999numerical}. For future reference, we let the rescaling be given by a diagonal matrix $\ma{D}\in\R^{n\times n}$. 

Assume that the output of the previous steps is given by $(\ve[k]{Y})_{k=0}^{N-1}$, where $\ma[k]{Y}=\ve{H}(\ma{D}\ve{X}(t_k))$, and that we are interested in modeling the evolution of $\ve[k]{Z}=\ve{G}(\ve[k]{Y})$, where both $\ve{H}:\R^n\rightarrow\R^q$ and $\ve{G}:\R^q\rightarrow\R^p$ are linear `permutation' maps that may exclude some elements. The maps are given the following interpretation: $\ma{H}$ selects the species that correspond to actual measurements, while the matrix $\ma{G}$ selects the species whose interdependencies we wish to infer. This allows us to remove species whose dynamics are faster than we can realistically sample, which behave as a constant with added white noise in steady state. Such species are detected by their time-series having a constant mean and approximately zero autocorrelation. In theory, a distinction is made between the cases of full state measurements for which good theoretical results exists and the case of hidden nodes which is more difficult \cite{gonccalves2008necessary}. For \emph{in vivo} experiments, the case of hidden nodes is prevalent. Indeed, the real \Ras/\cAMP/\PKA{} pathway is influenced by species which are not represented in Table \ref{tab:model} \cite{cazzaniga2008modeling,besozzi2012role}.

Let $\|\cdot\|_1:\R^{n\times n}\rightarrow[0,\infty)$ denote the entry-wise matrix $1$-norm given by $\|\ma{A}\|_1=\sum_{i,j}|\ma[ij]{A}|$, while $\|\cdot\|_2:\R^n\rightarrow[0,\infty)$ denote the Euclidean vector norm. The least absolute shrinkage and selection
operator (\LASSO) is an algorithm for solving sparse linear systems of equations and a key tool in compressive sensing. Using the model \eqref{eq:linear} to create an error to be minimized, the model is fitted to the data $(\ma{Z}(t_k))_{k=0}^{N-1}$ by solving \LASSO{} in the  Lagrangian form
\begin{align}
\tag{\textsc{lasso}}
\min_{\ma{B}\in\R^{p\times p }} \frac{1}{N}\sum_{k=0}^{N-1}\|\ma[k+1]{Z}-\Delta t\ma{B}\ma[k]{Z}\|^2_2+\lambda\|\ma{B}\|_1,\label{eq:lasso}
\end{align}
where the regularization parameter $\lambda\in[0,\infty)$  affects, roughly speaking, the trade-off between the goodness of fit and the sparsity of the regression parameters $\ma{B}\in\R^{p\times p}$. The matrix $\ma{B}$ is a submatrix of $\ma{A}$ in \eqref{eq:linear}, up to a change of basis. 
The $\frac1N$ and $\Delta t$ parameters are included to reduce the sensitivity of $\ma{B}$  to changes in the sample rate.

Consider that $M$ replicates of an experiment has yielded $M$ datasets $\mathcal{I}_i$, $i=1,\ldots M$, to be used for identification. For each $\mathcal{I}_i$, we infer a set of models $\ma{B}(\mathcal{I}_i,\lambda)$ using the \LASSO{} method for a range $[0,b]$ of values of $\lambda$. To determine the best value of the regularization parameter $\lambda$, we compare the ability of the models $\ma{B}(\mathcal{I}_i,\lambda)$ to predict the time-evolution of a validation data set $\mathcal{V}_{j(i)}$, $j(i)\in\{1,\ldots,K\}$, where $j(i)$ is selected at random. The validation data $\mathcal{V}_{j(i)}$ is the output of an experiment where the model organism is subjected to somewhat different conditions than for $\mathcal{I}_i$. For each set $\mathcal{I}_i$, we select the model that satisfies
\begin{align*}
\lambda=\argmin_{\mu\in[0,b]}\sum_{k=0}^{N-1}\|\ma[k+1]{Z}(\mathcal{V}_{j(i)})-\Delta t\ma{B}(\mathcal{I}_i,\mu)\ma[k]{Z}(\mathcal{V}_{j(i)})\|^2_2,
\end{align*}
where $\ma[k]{Z}(\mathcal{V}_{j(i)})$ is data from $\mathcal{V}_{j(i)}$. In an \emph{in vivo} setting, this approach corresponds to the common practice of a replicate experiment used to validate the original. Experiments that involve synchronization, in particular, should be repeated at least twice using different methods of synchronization since the process may induce artifacts in the cells \cite{futcher1999cell}.

\subsection{Modelling causal relations}
\label{sec:causal}

\noindent We wish to study causal relations in the \GRN. From the output of the \emph{in silico} experiment, all we know are changes in the molecule count numbers. A manipulation and invariance view of causality is hence appropriate: if, roughly speaking, after changing one gene we measure a change in the molecule count number of a protein, the gene is a direct or indirect cause of that change \cite{illari2014causality}. This idea is epitomized by the gene knock-out experiment, \ie the procedure of deactivating one or more genes at a time. However, such experiment designs suffer from a combinatorial explosion as we increase the number of genes to be manipulated, and does not account for redundancies in gene functionality \cite{illari2014causality}. As such, it is desirable to be able to reliably infer regulatory interactions from time-series data of \eg cell cycles rather than gene knockout experiments.

\begin{figure}[htb!]
	\centering
	\includegraphics[width=0.53\textwidth]{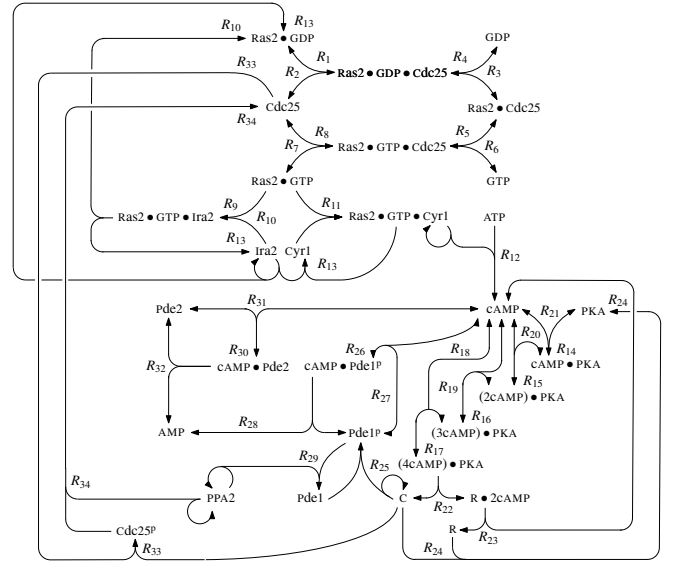}
	\caption{\label{fig:causal}Directed hypergraph $\mathcal{H}$ of the causal relations expressed by reactions $R_1$--$R_{34}$ in Table \ref{tab:model}. The hyperedges go from the reactants (no arrow) to the products (arrow). Hyperedges with arrows at both ends indicate that a reaction $R_i$ is reversed by another reaction $R_j$, for some $i,j\in\{1,\ldots,34\}$.}
\end{figure}

The causal relations underlying the reactions in Table \ref{tab:model} can be visualized using a hypergraph $\mathcal{H}$ where each reaction corresponds to a hyperedge, see Fig. \ref{fig:causal}. Note in particular that the graph is rather sparse, as is consistent with the assumption of Section \ref{sec:lasso}. To translate the ground truth into the modeling framework that we have adopted, \ie equation \eqref{eq:linear}, corresponds to converting the directed hypergraph in Fig.~\ref{fig:causal} into a directed graph with self-loops,
\begin{align}
\mathcal{D}=(\mathcal{V},\mathcal{F}),\label{eq:D}
\end{align}
where $\mathcal{V}=\{1,\ldots,30\}$ represents all the species in Table \ref{tab:model} and $\mathcal{F}=\cup_{i=1}^3\mathcal{A}_i$, where
\begin{align*} 
\mathcal{A}_1&=\{(i,j)\in\mathcal{V}\times\mathcal{V}\,|\,n_i S_i+\ldots\smash{\xrightarrow{c_k}} n_jP_j+\ldots,i\neq j\},\\
\mathcal{A}_2&=\left\{(i,j)\in\mathcal{V}\times\mathcal{V}\,|\,n_i S_i+n_jS_j\ldots\smash{\xrightarrow{c_k}} \sum_{l\neq i}n_lP_l\right\},\\
\mathcal{A}_3&=\{(i,i)\in\mathcal{V}\times\mathcal{V}\}.
\end{align*}
Each arc in $\mathcal{A}_1$ represents a reactant and a product, each arc in $\mathcal{A}_2$ two reactants of which at least one is consumed during the reaction, and each self-loop in $\mathcal{A}_3$ represent the fact that species which do not react persist existing. Note that one difference between the causality represented by $\mathcal{H}$ and $\mathcal{D}$: all species on the left-hand side of a reaction must be present for it to occur, but that requirement cannot be captured by a system of the form \eqref{eq:linear}. This would require \eqref{eq:linear} to include terms that are bilinear in the explanatory variables.

We adopt the following approach to approximately infer the \GRN{} topology. Given estimated values of the regression parameters $\ma{B}$, we assign a topology $\mathcal{G}(r)=(\mathcal{V},\mathcal{E}(r))$, where $\mathcal{U}=\{u_1,\ldots,u_q\}$ corresponds to the set of measured species, $\mathcal{V}=\{v_1,\ldots,v_p\}\subseteq\mathcal{U}$ is the set of species whose dynamics we wish to infer,  $\mathcal{E}(r)=\{(i,j)\in\mathcal{U}\times\mathcal{U}\,|\,|\ma[ij]{B}|\geq r\}$ are the causal relations, and $r\in[0,\max_{i,j}|B_{ij}|]$ is a threshold. By varying the threshold different causal models are obtained. The matrix $\ma{B}$ relate to $(\ve{X}(t_k))_{k=0}^{N-1}$ via the rescaling matrix $\ma{D}$ which is required for the optimization solver to converge. We could remove this dependence 
%
but it is our experience that the validation procedure gives a better result if we rescale $\mathcal{V}_{j(i)}$ (see Section \ref{sec:lasso}) rather than $\ma{B}$.

\subsection{Performance measure}
\label{sec:ROC}

\noindent To evaluate the performance of the network inference algorithm we focus on the relation of the inferred network topology to that of the ground truth $\mathcal{D}$ given by \eqref{eq:D}. We use a criteria known as the area under the precision-recall curve (\AUPR). Given an inferred representation of causal relations $G(r)$ and the ground truth $\mathcal{D}$, we can calculate the ratio of true positives to all estimated positives (precision, $|\{e\in\mathcal{E}(r)     \cap\mathcal{F}\}|/|\mathcal{E}(r)|$) and that of true positives to all positives (recall, $|\{e\in\mathcal{E}(r)\cap\mathcal{F}\}|/|\mathcal{F}|$). These are coordinates in \PR-space, \ie the unit square $[0,1]^2$ with precision on the ordinate and recall on the abscissa. By varying $r\in[0,\infty)$ we obtain a right to left curve from the point $(1,|\mathcal{F}|/|\mathcal{V}|^2)$ to some point in set $\{(0,s)\,|\,s\in[0,1]\}$. The area under this curve is the \AUPR. By plotting the \AUPR{} against the number of samples, we establish how the quality of inference depends on the temporal resolution of data, \ie the trade-off function. 

Let us make these notions more precise. A partition $\mathcal{P}=(t_k)_{k=0}^{N-1}$ of a time interval $[0,T]$ is a sequence of real numbers such that $t_0=0< t_1<\ldots<t_{N-1}=T$ \cite{abbott2001understanding}. Consider a number of partitions $\mathcal{P}_1,\ldots,\mathcal{P}_l$ of $[0,T]$ and the data corresponding to each partition $\mathcal{I}_i=(\ve{X}(t_k))_{t_k\in \mathcal{P}_j}$. The trade-off function is the discrete graph of the \AUPR{} obtained from inferring a model $\ma{B}(\mathcal{I}_j)$ which can be thresholded into a network $\mathcal{G}(r)$ over the sampling frequency $|\mathcal{P}_j|/T$. In this paper $T$ is constant, wherefore we plot the \AUPR{} against the number of samples $|\mathcal{P}_j|$. Although we define the trade-off function without specifying all details, it is clear that it depends on the \GRN{} inference method, in our case \LASSO.

Aside from the trade-off function that each experiment yields, we can consider a sample median trade-off function as the median over multiple experiments, and a true median trade-off function. The true trade-off function depends on the method used for inference. It is however clear that its value for zero samples is zero, and it seems likely that it converges to a constant in the limit of infinite samples although performance may deteriorate due to numerical reasons. If we know that to be the case, we can always prune samples and thereby reduce the sample rate to some practical value. As such, we expect the trade-off function to increase from 0 to some value in $[0,1]$ as $|\mathcal{P}_j|\rightarrow\infty$, or at least to increase in the case of sufficiently many samples.




Although the \AUPR{} is popular, it should be noted that there are other goodness of fit indices, \eg \ROC{} curves \cite{fawcett2006introduction}, or three-way \ROC{}s \cite{mossman1999three} and their respective integrals. We prefer the \AUPR{} since it is known to give a more realistic measure of performance than the \ROC{} when the distribution of positive and negative instances is heavily skewed \cite{saito2015precision}. This is the case for \GRN{} inference due to the sparseness of the network.  Random performance for the \AUPR{} is given by the number of true instances divided by the total number of instances, \ie $|\mathcal{F}|/|\mathcal{V}|^2$. An issue that benchmark and comparative studies face is that different methods are to some extent complimentary, and their ranking depends \eg on the type of network considered \cite{marbach2010revealing,marbach2012wisdom}. In this paper, we are interested in studying the performance of an algorithm relative to the quality of its input, \ie relative to itself. Fortunately, this relative performance should be less sensitive to the choice of inference algorithm, goodness of fit index, type of model, and type of network  than is the benchmark of one algorithm or comparative studies that benchmark multiple algorithms.



\section{Results}

\label{sec:results}

\begin{figure}[htb!]
	\centering
	\includegraphics[width=0.48\textwidth, clip=true, trim=25mm 10mm 20mm 10mm]{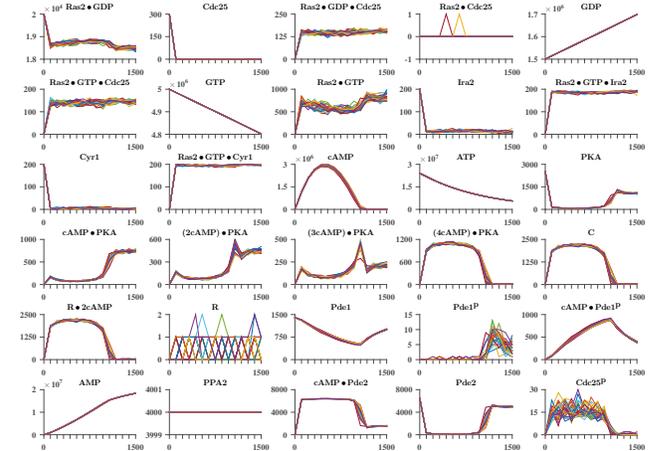}
	\caption{Twenty five draws from the solution to \eqref{eq:CME} for the reactions given by Table \ref{tab:model}--\ref{tab:copy} sampled 15 times uniformly over [0,1500]. \label{fig:15}}
\end{figure}
\begin{figure}
	\centering
	\includegraphics[width=0.48\textwidth, clip=true, trim=25mm 10mm 20mm 10mm]{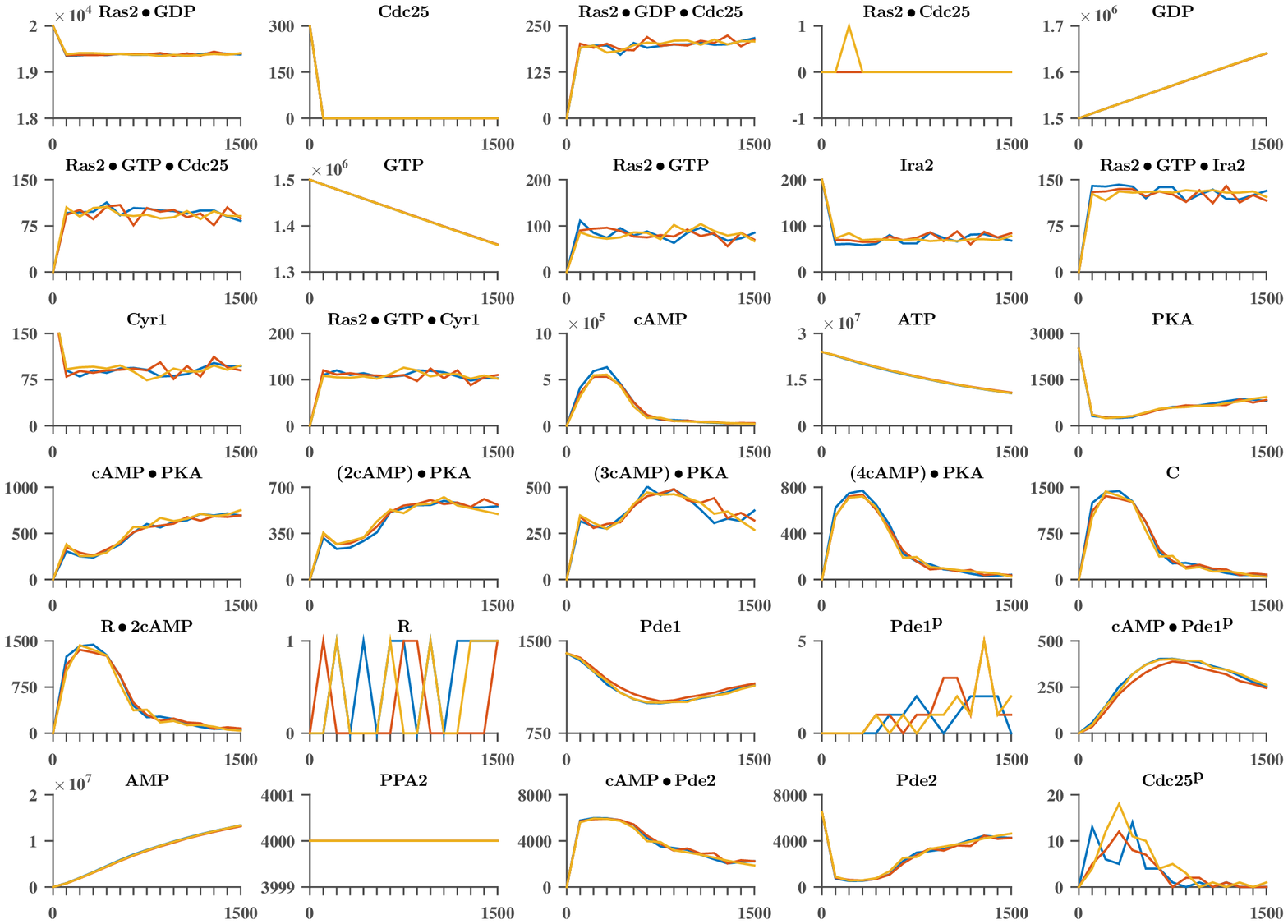}
	\caption{Three draws from the solution to \eqref{eq:CME} for the reactions given by Table \ref{tab:model}--\ref{tab:copy}, except  the initial value of \GTP{} is set to $\smash{1.5\cdot10^6}$, sampled 15 times uniformly over [0,1500]. \label{fig:val}}
	\includegraphics[width=0.5\textwidth, clip=true, trim=5mm 0mm 5mm 5mm]{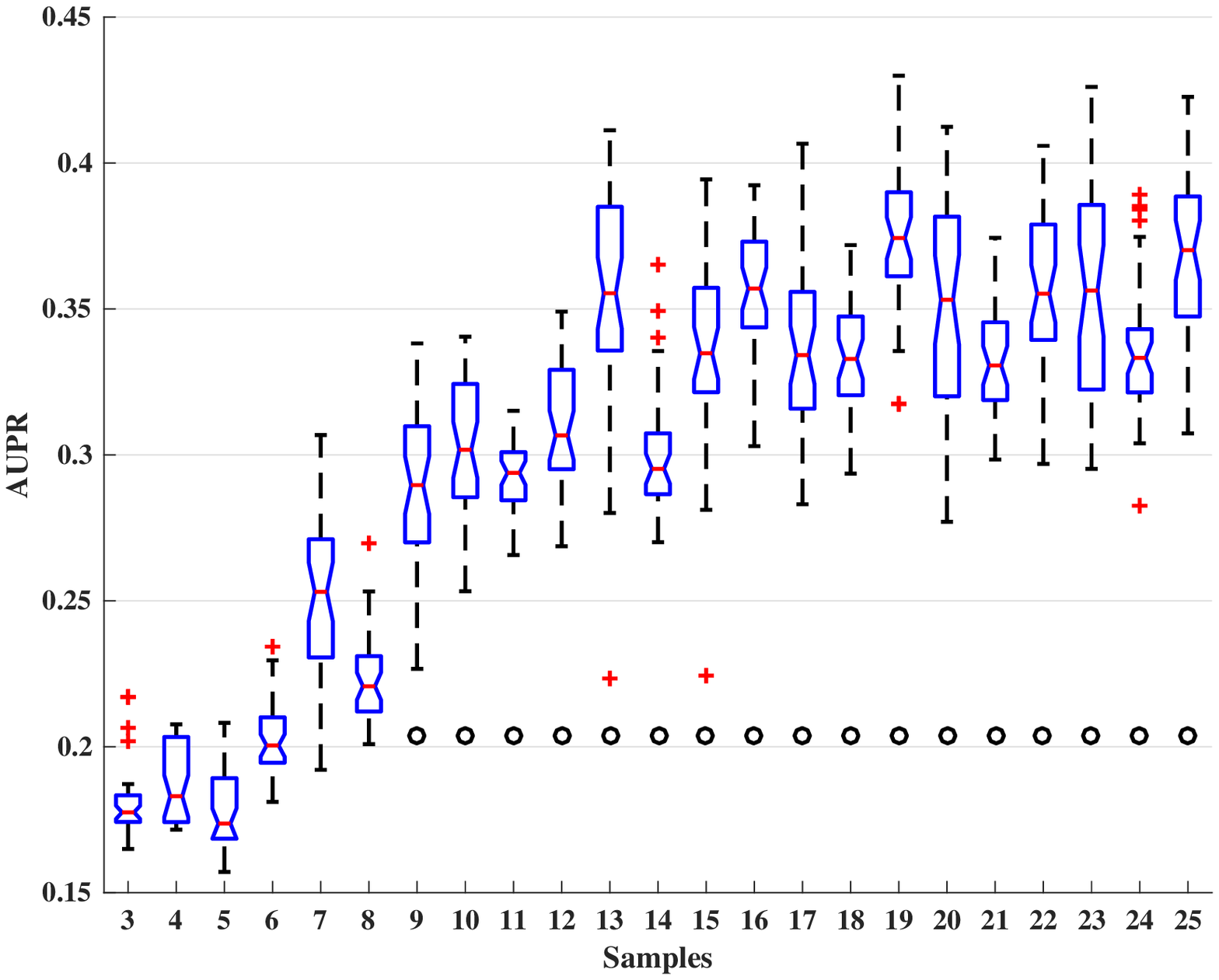}
	\caption{The trade-off function, \ie the \AUPR{} \vs the number of samples, based on 40 \emph{in silico} experiments. Each sample is represented by a boxplot: the waist is the median, the bottom and top edges of the box indicate the 25th and 75th percentiles respectively. The notches give a 95\% confidence interval for the true median. The whisker extend to the extreme data points besides outliers which are represented by plus signs (the whiskers enclose approximately 99.3\% of the data if it is normally distributed). The dots denote the performance of a random classifier, $\AUPR=|\mathcal{F}|/|\mathcal{V}|^2$.\label{fig:to}} 
	\vspace{3mm}
	\includegraphics[width=0.5\textwidth, clip=false, trim=0mm 5mm 0mm 5mm]{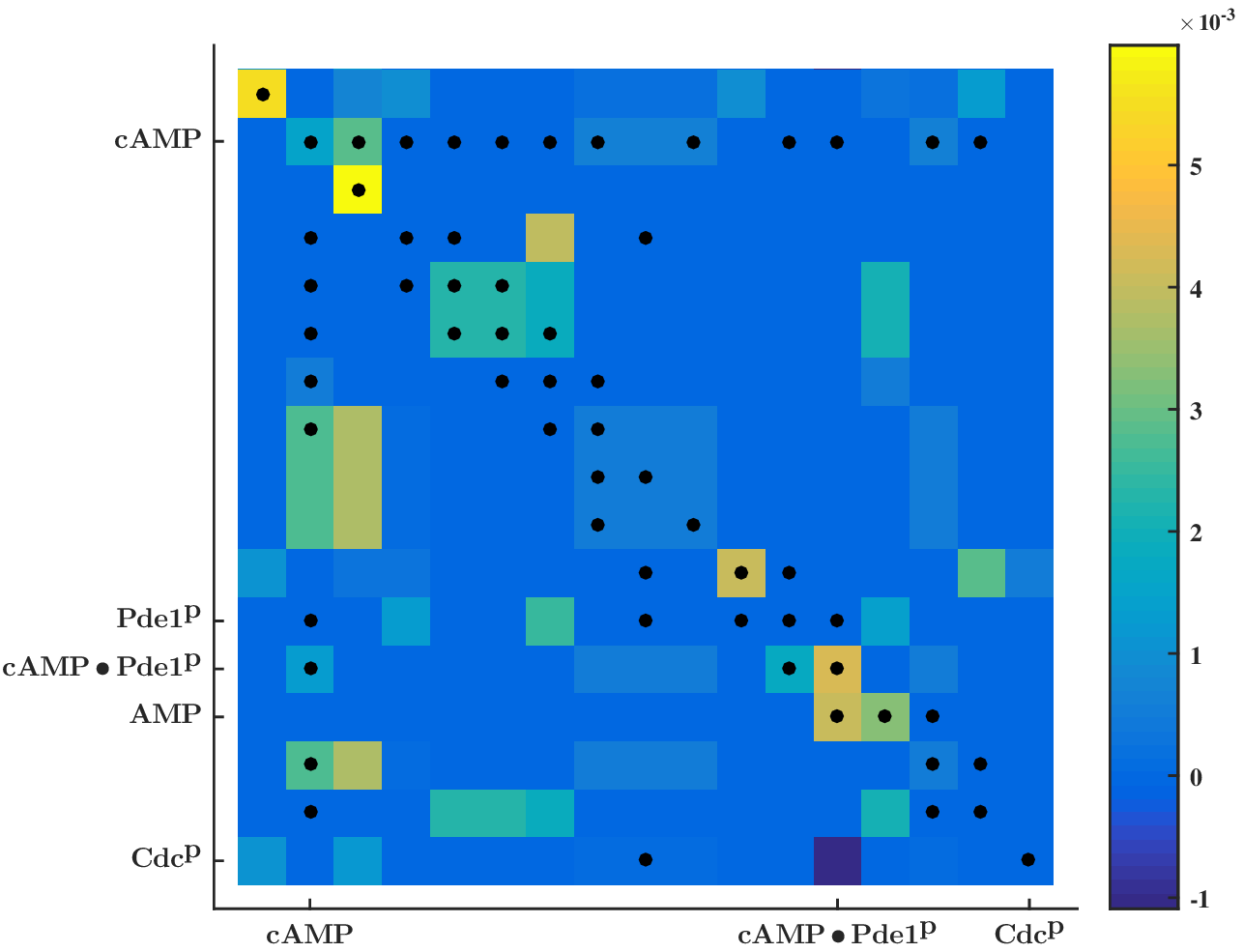}
	\caption{Heat map of $\ma{B}$ with the ground truth as black dots.\label{fig:B}} 
\end{figure}

\noindent We simulated 40 cells using the \ODM, each run encompassing $10^8$ reactions, resulting in datasets whose time span include $[0,3000]$. We keep the first 1500 time units, which correspond to 4.5--7.5 minutes \cite{cazzaniga2008modeling}. Realistically, this implies that we may sample 9--15 times at most (see Section \ref{sec:realistic}). The output of the simulation in the case of 15 samples is given in Fig. \ref{fig:15}. The intrinsic noise does not influence the overall shape of the trajectories, rather it is most pronounced in the species with low molecule count numbers such as \PdeO\highf{} and \CdcTF\highf. Fig \ref{fig:val} depicts a second set of 3 cells that is used as validation data (see Section \ref{sec:lasso}). The validation data is simulated from the glucose starved \SCer{} cell condition obtained by setting the initial value of the metabolite \GTP{} to $1.5\cdot10^6$ instead of $5\cdot10^6$ \cite{cazzaniga2008modeling}.

The species in the \CME{} model evolve over different time intervals, wherefore some are dormant or have already reached steady-state while others go through a transient state. This is typical of the \SCer{} cell cycle, where different genes are expressed during different phases. While the dense data $\smash{(X_i(\tau_k))_{k=0}^{10^8-1}}$ from the \SSA{} is not white noise on $[100,1500]$, the autocorrelation dissipate with time wherefore the sampled data $(X_i(t_k))_{k=1}^{N-1}$ on a time partition of length $N$ may be white noise. Species that are either white noise (\RasT\,\textbullet\,\GDP, \CdcTF, \RasT\,\textbullet\,\GDP\,\textbullet\,\CdcTF, \RasT\,\textbullet\,\GTP\,\textbullet\,\CdcTF, \RasT\,\textbullet\,\GTP, \IraT, \RasT\,\textbullet\,\GTP\,\textbullet\,\IraT, \CyrO,\, \RasT\,\textbullet\,\GTP\,\textbullet\,\CyrO, \textsc{r}), constant or practically constant after rescaling (\RasT\,\textbullet\,\CdcTF, \GDP, \GTP, \PPAT), on  $(t_k)_{k=1}^{N-1}$ are removed from the \GRN{} inference and evaluation process, compare with the 15 point time-series in Fig. \ref{fig:15}--\ref{fig:val}. It is possible to build a model of \eg \CdcTF{} given sufficently many samples from the interval $[0,100]$, but that would not be consistent with our assumption of slow sampling, \ie at most two samples per minute. 

Fig. \ref{fig:to} displays the trade-off function for the cases of 3--25 samples. The performance of a random classifier over this data yields an \AUROC{} of approximately 0.2. For the cases of 3--6 samples, we note that \LASSO{} performs on par with the random classifier. The performance in case of 7--15 samples is better than average with at least 95\% certainty (pointwise for each number of samples). Note that there is a trend of increasing performance with increasing samples. Cases of comparatively good or poor performance, like that of 7 and 14 samples respectively can partly be explained by variation in the data. Although not displayed in Fig. 4, more than 25 samples give diminishing returns with respect to the \AUROC. By identifying the true trade-off function with the sample medians, we could imagine that the shape of the trade-off function is approximately captured by a continuous sigmoid curve.

%

Consider the inclusion or exclusion of a phosphoproteomic study, \ie whether the species \PdeO\highf, \cAMP\,\textbullet\,\PdeO\highf, and \CdcTF\highf{} are measured or not. Fig. \ref{fig:to} is based on \emph{in silico} experiments that include phosphoproteomics. The regression parameters $\ma{B}$ of the best performing model with an \AUPR{} of $0.41$ is displayed in Fig. \ref{fig:B}. Note that neither \PdeO\highf{} could not be explained using the other data (last row have no true positives), nor is it helpful in explaining the other variables (last column is zero). The protein \PdeO\highf{} contributes a true positive (\cAMP\,\textbullet\,\PdeO\highf{} in its column) but it is mostly white noise followed by a short and noisy evolution. While the trajectory of \PdeO\highf{} is discernable in Fig. \ref{fig:15}, care must be taken as it becomes less so when the number of samples are reduced. However, \cAMP\,\textbullet\,\PdeO\highf{} is well explained with all positives identified on its row, and also manages to explain the evolution of \AMP{}, with two out of four true positives in its column. To have a true positive on the diagonal may not seem impressive, but it is valuable since it indicates that the model makes sense, \ie that it has some explanatory power aside from mere data fitting.



About 80\% of microarray time series in 2006 were short with lengths of 3--8 time points \cite{ernst2006stem}. For a study of the \Ras/\cAMP/\PKA{} pathway in \SCer{} where \GRN{} inference is done using the \LASSO{} method, such time-series would not suffice to infer the topology of the underlying network. It may still be possible to predict how the organism would react to changes in its environment, such as the difference between normal and low glucose levels as represented by the trajectories in Fig. \ref{fig:15} and Fig. \ref{fig:val} respectively. However, that model would not give us clues about the regulatory interactions inside the cell. In theory, it would be possible for an experimentalist that desires such an understanding to consult Fig. \ref{fig:to} and read off the minimum number of samples required to achieve a certain value of the \AUPR. In practice, the generality of our results need to be increased before it can become a useful tool in the laboratory.





\section{Discussion}

\label{sec:discussion}

\noindent This paper studies the trade-off between quality of inferred gene regulatory network models versus the temporal resolution of data in the case of full and partial state measurements corresponding to an experiment setup that either includes or excludes phosphoproteomics. The goodness of fit is characterized using the area under the curve of the precision-recall curve (\AUPR). In theory, experimentalists who desires a  particular \AUPR{} value may consult our graph of the trade-off function to see how many samples are needed to achieve that quality of inference. They can also determine if an increase in the number of samples, or the inclusion of phosphoproteomics, is worthwhile compared to their additional marginal and fixed experimental costs respectively. In practice, it is however clear that additional studies are needed before such a tool becomes mature enough to be of actual use in the laboratory. This paper should be considered as a proof-of-concept study. As such, its purpose is to establish a framework, showcasing how a study of the aforementioned trade-off can be conducted from simulation of data to the evaluation of an inference algorithm. 

\bibliographystyle{unsrt}
\bibliography{cdc2017}

%
%
%
%
%
%
%
%


%

\end{document}

%% file: myIncludes.tex
\usepackage{subdepth} 

\usepackage{graphics} 
\usepackage{epsfig} 
\usepackage{mathrsfs} 

\usepackage{amssymb, amsthm, amsfonts}  
\usepackage{amsmath,euscript,psfrag,color,graphicx}
\usepackage{graphicx}
\usepackage{latexsym}

\usepackage[small,bf,up,labelsep=period]{caption} 
\usepackage{quoting} 
\usepackage{lettrine} 
\usepackage{fmtcount}
\usepackage{calc} 
\usepackage{cite} 
\usepackage{multicol} 
\usepackage{centernot}
\usepackage{nicefrac}
\usepackage{dsfont}

\usepackage{lipsum}
\usepackage{psfrag}
\usepackage[export]{adjustbox} 

\PassOptionsToPackage{cmyk}{xcolor}
\usepackage[cmyk]{xcolor}
\selectcolormodel{cmyk}

\usepackage{bm}

\newcommand{\ve}[2][]{\ensuremath{\boldsymbol{\mathrm{#2}}}_{#1}}

\newcommand{\vh}[2][]{\ensuremath{\widehat{\boldsymbol{\mathrm{#2}}}_{#1}}}

\newcommand{\ma}[2][]{\ensuremath{\boldsymbol{\mathrm{#2}}}_{#1}}

\newcommand{\Prob}{\ensuremath{\mathds{P}}}

\newcommand{\R}{\ensuremath{\mathds{R}}}

\newcommand{\N}{\ensuremath{\mathds{N}}}

\newcommand{\ie}{\textit{i.e.}, }

\newcommand{\eg}{\textit{e.g.}, }

\newcommand{\etal}{\emph{et al.\ }}

\newcommand{\mtr}{\hspace{-0.3mm}\ensuremath{^\top}}

\newcommand{\diff}{\ensuremath{\mathrm{d}}}

\newcommand\raiseT[2]{\setbox0\hbox{$#1{#2}$}\raise\dp0\box0}

\newcommand{\vs}{\emph{vs.\ }}

\newcommand{\matlab}{\mbox{\textsc{matlab}}}

\DeclareMathOperator*{\argmin}{\text{argmin}}

\makeatletter
\newcommand{\raisemath}[1]{\mathpalette{\raisem@th{#1}}}
\newcommand{\raisem@th}[3]{\raisebox{#1}{$#2#3$}}
\makeatother

\definecolor{kthbluergb}{RGB}{25,84,166}
\definecolor{kthbluecmyk}{cmyk}{1,0.55,0,0}

\definecolor{kthblueA}{RGB}{25,84,166}
\definecolor{kthblueB}{RGB}{46,124,192}
\definecolor{kthblueC}{RGB}{112,153,209}
\definecolor{kthblueD}{RGB}{164,186,225}
\definecolor{kthblueE}{RGB}{211,220,241}

\newcommand{\highf}{\raise0.5ex\hbox{\scriptsize{p}}}

\newcommand{\LASSO}{\textsc{lasso}}

\newcommand{\CRN}{\textsc{crn}}

\newcommand{\GRN}{\textsc{grn}}

\newcommand{\SSA}{\textsc{ssa}}

\newcommand{\SCer}{\emph{S. cerevisiae}}

\newcommand{\SacCer}{\emph{Saccharomyces cerevisiae}}

\newcommand{\CME}{\textsc{cme}}

\newcommand{\FRM}{\textsc{frm}}

\newcommand{\DM}{\textsc{dm}}

\newcommand{\ODM}{\textsc{odm}}

\newcommand{\NRM}{\textsc{nrm}}

\newcommand{\ROC}{\textsc{roc}}

\newcommand{\AUROC}{\textsc{auroc}}

\newcommand{\PR}{\textsc{pr}}

\newcommand{\AUPR}{\textsc{aupr}}

\newcommand{\GDP}{\textsc{gdp}}

\newcommand{\CdcTF}{\textsc{c}dc{\footnotesize25}}

\newcommand{\RasT}{\textsc{r}as{\footnotesize2}}

\newcommand{\Ras}{\textsc{r}as}

\newcommand{\cAMP}{c\textsc{amp}}

\newcommand{\GTP}{\textsc{gtp}}

\newcommand{\AMP}{\textsc{amp}}

\newcommand{\PKA}{\textsc{pka}}

\newcommand{\NP}{\textsc{np}}

\newcommand{\CyrO}{\textsc{c}yr{\footnotesize1}}

\newcommand{\IraT}{\textsc{i}ra{\footnotesize2}}

\newcommand{\PPAT}{\textsc{ppa}{\footnotesize2}}

\newcommand{\PdeO}{\textsc{p}de{\footnotesize1}}

\newcommand{\sCYRO}{\textsc{c}yr{\scriptsize1}}

\newcommand{\sIraT}{\textsc{i}ra{\scriptsize2}}

\newcommand{\sPdeO}{\textsc{p}de{\scriptsize1}}

\newcommand{\sCdcTF}{\textsc{c}dc{\scriptsize25}}

\newcommand{\sPKA}{\textsc{pka}}

\newcommand{\sPPAT}{\textsc{ppa}{\scriptsize2}}

\newcommand{\sPdeT}{\textsc{p}de{\scriptsize2}}

\newcommand{\sRasT}{\textsc{r}as{\scriptsize2}}

\newcommand{\sGDP}{\textsc{gdp}}

\newcommand{\sGTP}{\textsc{gtp}}

\newcommand{\sATP}{\textsc{atp}}

\newcommand{\scAMP}{c\textsc{amp}}

\newcommand{\sR}{\textsc{r}}

\newcommand{\sC}{\textsc{c}}

\newcommand{\sAMP}{\textsc{amp}}

\newcommand{\shighf}{\raise0.7ex\hbox{\tiny p}}

\newcommand{\mytextbullet}{\,\textbullet\,}

\newcommand{\ODE}{\textsc{ode}}

\newtheoremstyle{myplain} 
{\topsep}                    
{\topsep}                    
{\itshape}                   
{}                           
{\bfseries\sffamily}         
{.}                          
{.5em}                       
{}  

\theoremstyle{myplain}

\newtheoremstyle{mydefinition} 
{\topsep}                    
{\topsep}                    
{}                   
{}                           
{\bfseries\sffamily}         
{.}                          
{.5em}                       
{}  

\newtheoremstyle{myremark} 
{\topsep}                    
{\topsep}                    
{}                   
{}                           
{\itshape}         
{.}                          
{.5em}                       
{\thmname{#1}\thmnumber{ #2}\thmnote{.#3}}  

\theoremstyle{myremark}


\newcounter{parentnumber}

\newcommand\Tstrut{\rule{0pt}{2.6ex}}         
\newcommand\Bstrut{\rule[-0.9ex]{0pt}{0pt}}   